\documentclass{iaus}
\usepackage{graphicx,amsmath}
\newcommand{\EQ}{\begin{equation}}
\newcommand{\EN}{\end{equation}}
\newcommand{\EQA}{\begin{eqnarray}}
\newcommand{\ENA}{\end{eqnarray}}

\newcommand{\mean}[1]{\overline #1}

{}
{}
{}

{}
{}
{}
{}
{}
{}
{}
{}
{}
{}
{}
{}
{}
{}
{}
{}
{}
{}
{}
{}

{}
{}
{}

{}

{}
{}

\newcommand{\UU}{\mbox{\boldmath $U$} {}}

\newcommand{\erf}{{\rm erf}}

\def\etaT{\eta_{\rm T}}

\newcommand{\diff}{\rm{cm}^2\!/s}
\graphicspath{{./figs/}{./png/}}

\title[Magnetic feature tracking speed]  
{Magnetic feature tracking, what determines the speed?}

\author[G. Guerrero, M. Rheinhardt \& M. Dikpati]    
{Gustavo Guerrero$^{1,3}$, Matthias Rheinhardt$^{2,3}$, 
  \& Mausumi Dikpati$^4$ }

\affiliation{
$^1$Solar Physics, HEPL, Stanford University, Stanford, CA, 94305-4085, USA 
\\email: {\tt gag@stanford.edu}\\[\affilskip]
$^2$Department of Physics, FI-00014 University of Helsinki, Finland\\  
email: {\tt rheinhar@mappi.helsinki.fi} \\[\affilskip]
$^3$NORDITA, Roslagstullsbacken 17, Stockholm, Sweden \\
$^4$High Altitude Observatory, NCAR, Boulder, CO, 80301, USA\\
email:  {\tt dikpati@ucar.edu}
}
\pubyear{2011}
\volume{286}  
\pagerange{119--126}
\jname{Comparative Magnetic Minima: Characterizing quiet times in
  the Sun and stars}
\editors{A.C. Editor, B.D. Editor \& C.E. Editor, eds.}
\begin{document}

\maketitle

\begin{abstract}
Recent observations revealed that small magnetic elements
abundant at the solar surface move poleward with a velocity 
which seems to be lower than the plasma velocity $\UU$. 
\cite[Guerrero et al. (2011)]{guerr+etal_11} explained this discrepancy 
as a consequence of diffusive spreading of
the magnetic elements due to a  positive radial gradient 
of $|U_\theta|$.  As the gradient's sign (inferred by local 
helioseismology) is still unclear, cases with a negative gradient are 
studied in this paper. Under this condition,
the velocity of the magnetic tracers turns out to 
be larger than the
plasma velocity, in disagreement with
the observations. Alternative mechanisms for explaining
them independently are proposed. For the turbulent magnetic pumping 
it is shown that it has to be unrealistically strong to reconcile 
the model with the observations. 
\keywords{Sun: activity, Sun: magnetic fields}
\end{abstract}

\firstsection 
\section{Introduction}

Amongst the axisymmetric constituents of the plasma flow in the Sun's 
convection zone, the meridional circulation (MC), although being much 
slower than the differential rotation, is gaining growing interest which
arises from its potential importance for the solar dynamo.
Certainly any advanced non-linear mean-field model of this dynamo 
will have to include the MC
to cover the full interaction between
mean field and motion.
In particular however, flux-transport dynamo models 
depend already on the {\em kinematic level} crucially on the MC as 
it is the ingredient which allows to close the dynamo cycle.
At the same time it explains naturally both
the equatorward migration of active regions
and the poleward drift of weak poloidal field
elements during the solar cycle.

The MC in the Sun is accessible to a variety of  techniques 
out of which Doppler measurements
and helioseismological time-distance or ring diagram analyses
tend to agree in the surface speed
profiles with a peak of $\approx20$m/s 
at a latitude $\approx35^{\circ}$.  However, attempts to 
measuring the MC
velocity using small magnetic structures as tracers of the flow 
(\cite[Komm et al., 1993, Hathaway \& Rightmire, 2010]
     {komm+etal_93,hatha+righ_10}) reveal systematic differences from 
these results.
Compared to Doppler measurements (\cite[Ulrich 2010]{ulrich2010})
magnetic feature tracking (MFT) speeds
can be as much as $\approx 30$\% smaller.
This indicates that the magnetic field is {\it not completely frozen} 
in the plasma, which is reasonable given the
assumed high turbulent magnetic diffusivity $\etaT$. Due to
diffusive spreading of the field, its
advection is affected by the depth 
dependence of the MC velocity.
On top of advection and diffusion,
turbulent motions may be contributing by
an extra advection term due to turbulent magnetic pumping 
or by the more complicated interaction with  dynamo waves.

In a previous work (\cite[Guerrero et al., 2011]{guerr+etal_11}, hereafter
GRBD) we studied numerically the
kinematic evolution of small bipolar magnetic elements at 
different latitudes in the northern hemisphere. 
For simplicity we considered turbulent diffusion
and advection due to a predefined MC
as the only magnetic transport mechanisms. In a 1D
(i.e., surface) version of the model, the
northern spot acquires an effective velocity which is higher
than the flow speed, but the southern
one travels slower than the fluid.  
However, on average,
or considering the center of the bipolar region, its
speed clearly  fits the fluid speed. 
This result is independent of the value of $\etaT$.
In the 2D model, on the other hand, we found that the MFT speed 
differs from the fluid velocity provided
that the frozen-in condition is not satisfied.
Moreover, we demonstrated that the
relevant threshold value of $\etaT$
depends on the radial gradient of 
the latitudinal velocity, $\partial_r |U_{\theta}|$,
assumed positive. The difference between MFT and flow speeds 
increases with $\partial_r |U_{\theta}|$.
Thus, this simple model is able to explain the discrepancy 
between the Doppler and MFT observations.

Modern local helioseismological techniques
allow also to infer the radial variation of the MC
speed. Unfortunately, time-distance and 
ring-diagram analyses have not given consistent results.  
The first method yields that at the first few
megameters beneath 
the photosphere the latitudinal velocity decreases inwards, 
i.e., a positive $\partial_r |U_{\theta}|$ 
(\cite[Zhao et al. 2011]{zhao+etal_11}).
By contrast, the results from the second method indicate a
slight inward {\em increase}, i.e., a negative gradient
albeit restricted to middle latitudes. 
(\cite[Gonz\'{a}les Hern\'{a}ndez et al. 2006]{gh+etal_06}). 
Such velocity profiles were not considered in GRBD and
the aim of this paper is to fill this gap as well as to
discuss other effects that may influence MFT speeds.

\section{MFT speed with a negative radial gradient of $\,|U_{\theta}|$}
The method followed here to simulate 
the evolution of small bipolar magnetic structures is described 
in detail in GRBD. We also use the same notation.
Briefly, we solve the induction equation
in axisymmetry for the azimuthal component of the vector potential, 
$A_{\phi}$ in the variables $r$ and $\theta$. The prescribed  meridional 
flow fed into  the model is given by eqs. (4-6) of GRBD
and has a single cell per hemisphere. In the present paper
$\etaT$ is assumed to be
constant across the domain. Bipolar regions
(with spot separation ~3$^{\circ}$ and depth $0.04R$) at  
20 different latitudes from the pole to equator are
evolved over a time  interval of two weeks. Their 
velocity is computed as an average over this time span
through the differences in final and initial latitudes 
(see eq. (8) of GRBD). 
\begin{figure}
\centering
\vspace{-2mm}
\includegraphics[width=0.49\columnwidth]{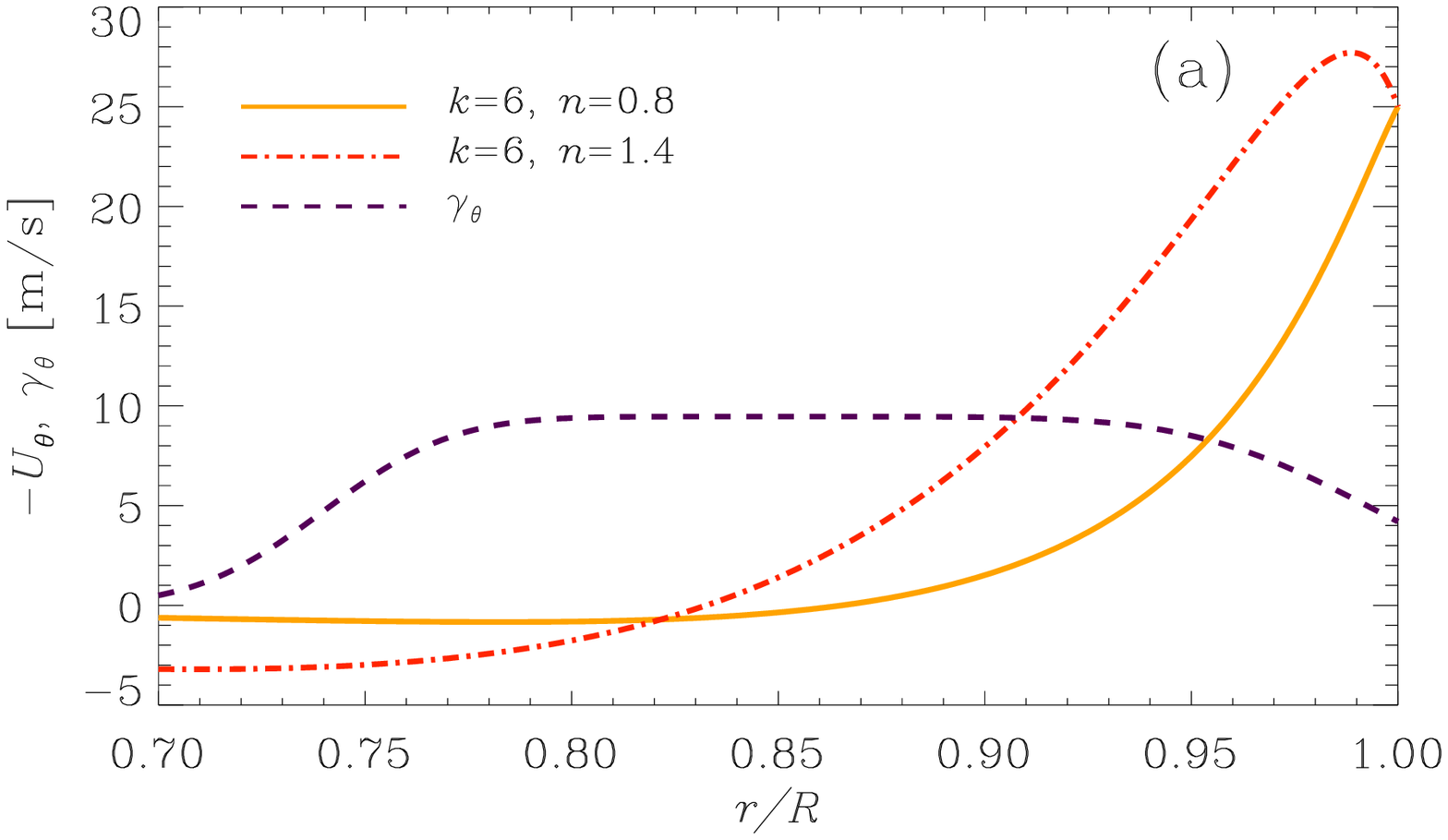}
\includegraphics[width=0.49\columnwidth]{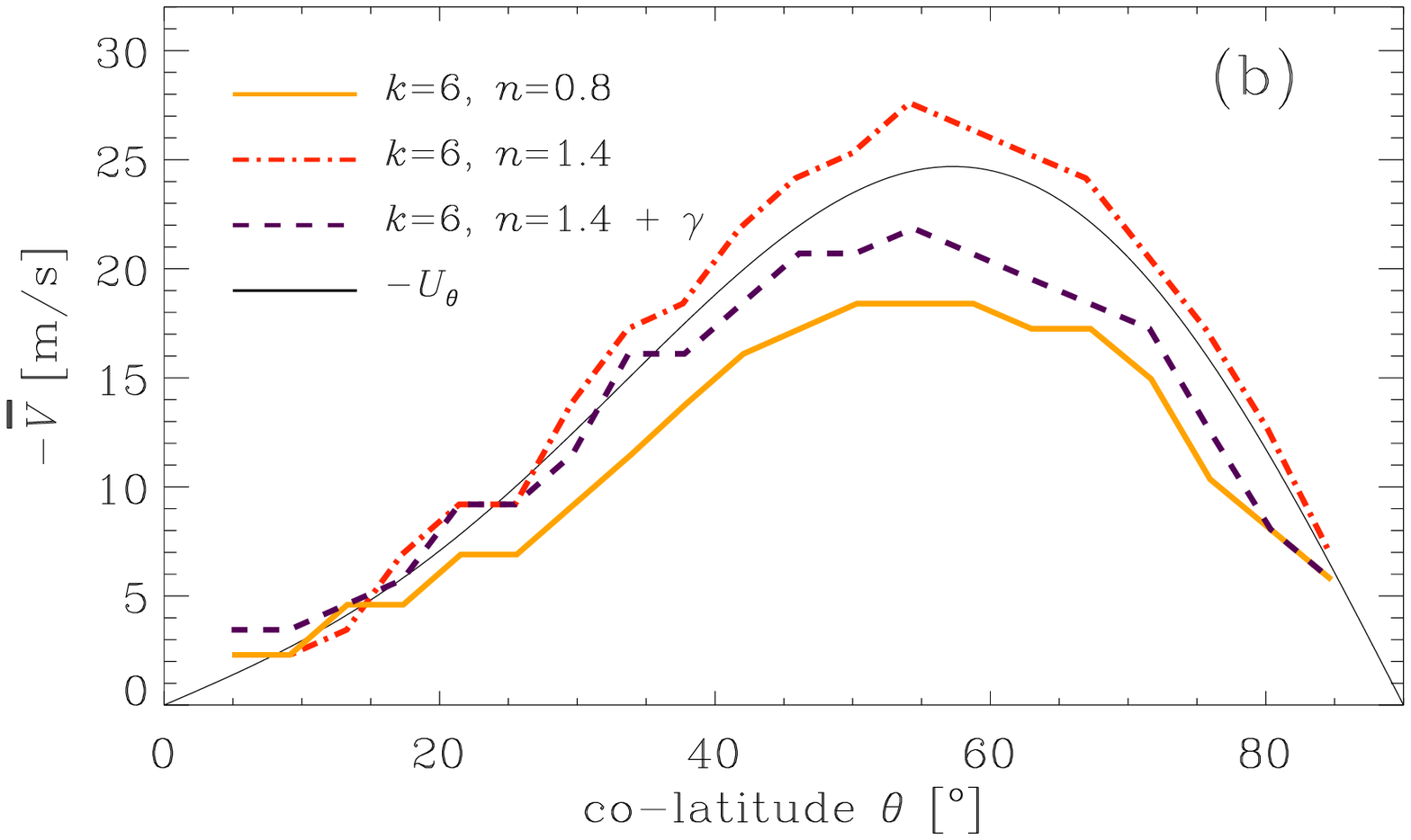}
\caption{
a) Radial profile of $U_\theta(r,57^\circ)$ for different 
values of $n$. Dashed line: radial profile
of $\gamma_{\theta}$, see eq. \eqref{eq:thp}.\; b) Corresponding 
tracking velocities,  $\mean{V}$, 
for three cases: $\partial_r |U_{\theta}|>0$ (yellow/continuous) and  
$\partial_r |U_{\theta}|<0$ with 
(dashed) and without (red/dot-dashed) 
pumping. $\etaT=10^{12}\diff$ throughout. }
\label{fig:uthrad}
\end{figure}
To obtain a negative radial gradient of $|U_{\theta}|$ we 
preserve its general mathematical form (eq. (5) of GRBD), but choose 
the parameters of this ansatz as  $k=6$ and $n=1.4$
which results in a negative $\partial_r |U_{\theta}|$ within a shallow 
surface-near layer of roughly 2\% of the thickness of the convection zone 
(see Fig. \ref{fig:uthrad}a).   It is worth mentioning
that the stress-free boundary condition
cannot be satisfied  with a negative gradient.
With this $U_\theta$ profile the velocities of the magnetic tracers
turn out to be larger than the flow velocity
(red/dot-dashed line in Fig. \ref{fig:uthrad}b). This result
does not come unexpected since the field 
lines underneath the surface are pushed northwards by a
flow faster than the surface one. By magnetic tension
the field lines at the surface are then also accelerated.
A snapshot of the magnetic field after two weeks of evolution
is shown in Fig. \ref{fig:brpos}a.

\section{Additional transport mechanisms}
Given that a negative  $\partial_r |U_{\theta}|$ does not allow to
reproduce the observations of {\em decelerated} magnetic tracer motion 
at the surface, one could be tempted to argue that
the results of the ring diagram analysis could be incorrect.
This conclusion, however, would be premature as our model is by 
no means complete. As mentioned in GRBD,  studying axisymmetric, 
i.e., averaged fields requires the inclusion of the
full mean electromotive force, from which only the turbulent diffusion term
has so far been considered.

What about other constituents?
Under anisotropic conditions, turbulence acts like a mean advective motion,
called {\em turbulent magnetic pumping} and is described by a
vector $\boldsymbol{\gamma}$. In the northern hemisphere,
its $\theta$-component is found to be positive, 
i.e., equatorwards whereas its radial component is mainly 
negative (downwards), see \cite{KKOS06}.
Being helical, convective turbulence in the rotating Sun exhibits
also an $\alpha$-effect which together with
differential rotation in the uppermost 35Mm of the convection 
zone (the so called near-surface shear layer) may well
enable  a turbulent dynamo 
of the $\alpha\Omega$ type. 
As  $\partial_r \Omega < 0$  there and $\alpha$ is expected to 
be positive in the northern hemisphere,
the dynamo wave should, according to the Parker-Yoshimura rule, 
propagate equatorwards,  contrary to the MC.
This scenario is difficult to study
as it requires establishing a full nonlinear mean-field dynamo model of 
the surface shear layer.  Turbulent pumping,  in contrast,  
can easily be introduced as an additive contribution to the  MC.
\begin{figure}
\centering
\vspace{-2mm}
\includegraphics[width=0.7\columnwidth]{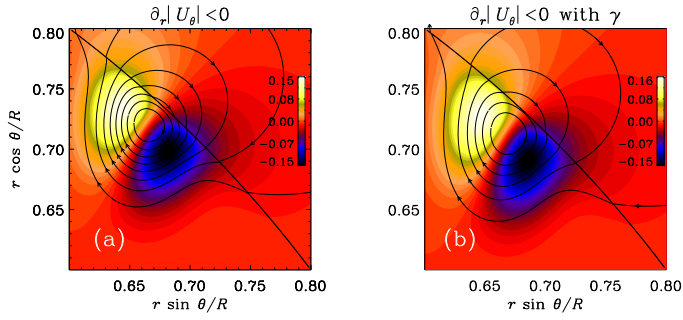}
\caption{Magnetic field lines and $r$ component (color coded) after 
two weeks propagation, $\etaT=10^{12}\diff$.
Velocity profile as in Fig.~\ref{fig:uthrad}, red curve.
a) without, b) with pumping, $\gamma_{0 \theta}=10\,\rm{m}/\rm{s}$, 
see eq.\eqref{eq:thp}}.
\label{fig:brpos}

\vspace{-5mm}
\end{figure}
We do so employing profiles for $\boldsymbol{\gamma}$
adapted from \cite{GDP08} (hereafter GDP08):   
\begin{equation}
\hspace*{-4mm}\begin{alignedat}{3}
  \gamma_{\theta} &=  &&\gamma_{0 \theta}\left[1+\erf \!
    \left(\frac{r-0.74}{0.035}\right) \right]& \!\!&\left[1-\erf \!
    \left(\frac{r-0.995}{0.05}\right) \right] 
   \cos\theta \sin^4\theta, \quad \quad \quad \quad \\   
  \gamma_{r} &= &\,-&\gamma_{0 r}\left[1+\erf\!            
    \left(\frac{r-0.7}{0.015}\right) \;\:\right]& \!\!&\left[1-\erf\!
    \left(\frac{r-0.975}{0.1}\right) \right] \!\!
    \left[\exp\! \left( \frac{r-0.715}{0.25} \right)^2
    \!\!\!\cos\theta +1 \right] \!.
\end{alignedat}\hspace*{-4mm}\label{eq:thp}
\end{equation}
Similar to GDP08, we set $\gamma_{0 r}$ and $\gamma_{0 \theta}$ 
such that the maximum of  $\gamma_{\theta}$ is 2.5 times larger
than that of $\gamma_{r}$.
\footnote{ 
Note that $\gamma_\theta$ slightly differs from
GDP08 insofar  it is here $\ne 0$ at the surface. 
In the underlying DNSs a 
$\gamma_{\theta}$ vanishing at the surface could be 
an artifact due to the boundary conditions.  
}
The radial profile of $\gamma_{\theta}$, which is the 
more relevant component for this study, is shown with a broken 
line in Fig. \ref{fig:uthrad}a. 
From direct numerical simulations  (DNS) \cite{KKOS06}
have estimated that the maximum 
$\gamma_\theta$  should  be $\approx2.5$m/s. According
to eq. \eqref{eq:thp} its value
at the surface would then be $1.11$m/s. 
Given that the surface amplitude of $|U_\theta|\approx 25$m/s,
this small value cannot significantly modify
the results found in the previous section. 
Hence we have increased $\gamma_{0 \theta}$ progressively and found that an
MFT speed profile that roughly coincides with
that of the flow speed is obtained with $\gamma_{0 \theta}=5 \rm{m/s}$ 
(i.e., $2.2$m/s at  the surface or 
$\approx10$\% of the flow speed). 
For larger values of $\gamma_{0 \theta}$ the MFT speed
is smaller than the flow speed.
In Fig. \ref{fig:uthrad}b we show the result obtained 
with the extraordinarily high value $\gamma_{0 \theta}=10$m/s 
(see broken lines). 
Note that the latitudinal profile of $\gamma_{\theta}$ 
(eq. \eqref{eq:thp}) peaks at 63$^{\circ}$ while the flow profile 
does at 57$^{\circ}$. This is
interesting because the resulting profile of the MFT speed peaks 
at a  different latitude than the flow,
just as it is obtained in observations (see fig. 10 of 
\cite[Ulrich, 2010]{ulrich2010}). 
Fig. \ref{fig:brpos} compares the morphology of the magnetic field from
the simulation with pumping (b) to that without (a). 
Note that the magnetic field is transported further to the north
in the case without magnetic pumping

\begin{discussion}
Negative gradients $\partial_r |U_\theta|$ at the solar surface, 
as suggested by helioseismological ring-diagram analyses,
result in MFT speeds higher than the plasma flow speed. 
By enhancing our flux transport model with turbulent magnetic pumping,
we tried to reconcile the helioseismological results with the actually 
oberserved {\em lower} MFT speeds.
However, high amplitudes of the pumping velocity not supported by 
DNS are needed. 
We have to conclude that either: (i) the negative gradient of 
$|U_\theta|$, suggested by  ring-diagram analysis, if real, 
should be smaller than considered here, (ii) 
the pumping velocities from DNS are unrealistically low, 
which cannot be excluded because the setup in  \cite{KKOS06} differs 
from the solar conditions, or (iii) the inclusion of the $\alpha$ effect 
in the flux transport model,   enabling a dynamo process in the Sun's 
surface shear layer,  is crucial. For the latter,
we have to think of a short-wavelength dynamo wave traveling equatorwards. 
Magnetic quenching would cause a corresponding
$\theta$ modulation of $\alpha$,  $\boldsymbol{\gamma}$ 
and $\etaT$ whereas the mean Lorentz force would modulate  
$U_\theta$ and $\Omega$.
Here, quenching of $\etaT$ is perhaps most promising as 
the value of $\etaT$  ``decides''to what degree the field 
is frozen in the fluid. 
\end{discussion}
\vspace{2mm}
\noindent
\textit{Acknowledgment}: We 
 thank A. Mu\~{n}oz-Jaramillo for pointing out the
 negative radial gradient of $|U_{\theta}|$ in the ring-diagram analysis and
 A. Kosovichev for his valuable comments.

\end{document}